\documentstyle[12pt,epsf]{article}

\begin{document}

\begin{center} 

{\Large \bf Dark energy from the Trans-Planckian
physics}\footnote{Talk given at CICHEP-01, Cairo, Jan. 2001} \\ 
\end{center}


\begin{center}
L. Mersini \\

{\small \it Scuola Normale Superiore \\ Pzza. dei Cavalieri n. 7, 56126-Pisa (Italy)  }
\end{center}







{\small 
As yet, there is no underlying fundamental theory for the
transplanckian regime. There is a need to address the issue of how the
observables in our present Universe are affected by processes that may
have occurred at superplanckian energies (referred to as the
{\it transplanckian regime}). Specifically, we focus on the impact the 
transplanckian regime has on the dark energy.  
We model the transplanckian regime by introducing
a 1-parameter family of smooth non-linear 
dispersion relations which modify the frequencies at very short
distances.
A particular feature of the family of dispersion functions chosen  is
the production of 
ultralow frequencies at very high momenta $k$ (for $k>M_P$). 
We show that the range of modes with frequencies equal or less than
the current Hubble rate $H_0$ which are still frozen today 
provides a strong candidate for the  {\it dark energy}
of the Universe.
}

\section{Introduction} 

There is strong evidence our Universe is accelerating due to a nearly
constant vacuum energy, the dark energy, that accounts for about 70\%
of the present total energy density. 
Although the SN1a data indicating acceleration is inconclusive, the
observed CMB spectrum ($\Omega_0\simeq 1.1 \pm 0.07$) as well as the
measurements of the Hubble constant (i.e. the age of the Universe)
strongly support the case of a universe with a large dark energy
component \cite{cmb}. 
There are many efforts in 4 or higher dimensional models that address
the issue of producing such a small vacuum energy  from the cosmological fundamental mass scale. An equally
challenging question is
why the  dark energy is the dominant component today . From the particle physics point of view, the main task is to
explain why its value is $10^{-122}$ in units of the Planck mass
$M_P$. Generally in these efforts, either a large amount of fine-tuning, or the
introduction of some ad-hoc, unnaturally small parameter is required. 

We propose in this talk a new approach for the origin of dark
energy, which is based on contributions from the transplanckian regime. This alternative model makes use of the `freeze-out by expansion` mechanism of
a range of  modes in the transplanckian regime
that contribute the correct amount to the dark energy of the
universe, without any fine-tuning and with 
$M_{pl}$ as the fundamental  scale of the theory~\cite{mar}.

In an expanding universe, the physical momenta are blueshifted back in
time; therefore some of the observed low values of the momenta today that
contribute to the CMBR spectrum may have originated from the
transplanckian regime in the Early Universe. Similar issues apply to
the Hawking radiation in Black Hole physics.
In a series of papers~\cite{jacobson}, it was demonstrated that the
Hawking radiation remains 
unaffected by modifications of the ultra high energy regime, expressed
through the modification of the usual linear dispersion relation at
energies larger than a certain ultraviolet scale $k_C$. Following a
similar procedure, in the case of an expanding
Friedmann-Lemaitre-Robertson-Walker (FLRW) spacetime, 
it has been showed~\cite{brand,mar} that the primordial
scale invariant power spectrum 
can be recovered if the dispersion relation for the metric
perturbations is a smooth function such
that it gives rise to an  adiabatic time-evolution of the modes. 

 We incorporate the concept
of superstring duality (which applies at transplanckian regimes)  in
the family of dispersion models for the transplanckian regime below \cite{mar} by
choosing a particular family of dispersion relations that exhibits
{\it dual} behavior\footnote{For example, when compactifying superstring
theory in a torus topology, of large radius $R$ and winding radius $r$, the
frequency mode spectrum is dual in the sense that $R$ and $r$ are related as
$r=1/R$. This means that each normal mode with a frequency $n/R$,
where $n$ is an integer, 
has its dual winding mode with decreasing energy that goes like
$1/r=R$~\cite{superstrings}.}, 
 i.e. appearance of ultra-low mode frequencies
both at low and high momenta $k$.

\section{Dark Energy from the ``Tail''}
Let us start with the generalized Friedmann-Lemaitre-Robertson-Walker
(FLRW) line-element which, in the presence of scalar and tensor
perturbations, takes the form \cite{equations}
\begin{eqnarray}
ds^2&=& a^2(\eta) \left\{ -d\eta^2 + \left[ \delta_{ij} + h(\eta,{\bf
n})Q \delta_{ij} \right. \right.  \nonumber \\ & &\left. \left. + h_l
(\eta,{\bf n}) \frac{Q_{ij}}{n^2} + h_{gw}(\eta,{\bf n}) Q_{ij}
\right] d x^i d x^j \right\} \,,
\label{frw}
\end{eqnarray}
where $\eta$ is the conformal time and $a(\eta)$ the scale factor. The
dimensionless quantity ${\bf n}$ is the comoving wavevector, related
to the physical vector ${\bf k}$ by ${\bf k}= {\bf n}/a(\eta)$ as
usual. The functions $h$ and $h_l$ represent the scalar sector of
perturbations while $h_{gw}$ represents the gravitational waves.
$Q(x^i)$ and $Q_{ij}(x^i)$ are the eigenfunction and eigentensor,
respectively, of the Laplace operator on the flat spacelike
hypersurfaces. For simplicity, we will take a scale factor $a(\eta)$
given by a power law,
$a(\eta)=|\eta_c/\eta|^{\beta}$, where $\beta \geq 1$ and $|\eta_c|=
\beta/H(\eta_c)$. The initial
power spectrum of the perturbations can be computed once we solve the
time-dependent equations in the scalar and tensor sector. The mode
equations for both sectors reduce \cite{equations} to a
Klein-Gordon equation of the form
\begin{equation}  
\mu_n^{\prime \prime} + \left[ n^2 - \frac{a^{\prime \prime}}{a} \right]
\mu_n=0 \,,
\label{kg}
\end{equation}
where the prime denotes derivative with respect to conformal time.
Therefore, studying perturbations in a FLRW background is equivalent
to solving the mode equations for a scalar field $\mu$ related
to the perturbation field  in the expanding background. 

In what follows, we replace the linear relation $\omega^2(k) = k^2
=n^2/a(\eta)^{2}$ with a nonlinear dispersion relation (which is the
class of Epstein functions)
$\omega(k)=F(k)$, such that 
\begin{equation}
n_{eff}^2 = a(\eta)^2 F(k)^2 = n^2 \left(\frac{\epsilon_1}{1+ e^x} +
\frac{\epsilon_2 e^x}{1+e^x} + \frac{\epsilon_3 e^x}{(1+e^x)^2}\right) \,,
\label{neff}
\end{equation}
where $x=(k/k_C)^{1/\beta}=  A |\eta|$, with $
A=(1/|\eta_c|)(n/k_C)^{1/\beta}$.
This dispersion function encapsulates the T-duality behaviour.Then, the equation for the scalar and tensor
perturbations, that we need to solve, takes the form
\begin{equation}
\mu_n^{\prime \prime} + \left[n^2_{eff} - (1- 6\xi)\frac{a^{\prime
\prime}}{a}\right] \mu_n = 0 \,.
\label{modeneff}
\end{equation}
This equation represents particle production in a time-dependent
background. We use the method of Bogolubov transformation to calculate
the spectrum of particles created and their energy. Although at late
time the background is asymptotically flat, the wavefunction will be in 
a squeezed state due to the curved background that mixes positive and
negative frequencies. The evolution of the mode function $\mu_n$ at
late times fixes the Bogolubov coefficients $\alpha_n$ and $\beta_n$,
\begin{equation}
\mu_n \rightarrow_{\eta \rightarrow +\infty} \frac{\alpha_n}{\sqrt{2
\Omega^{out}_n}} e^{-i\Omega^{out}_n \eta} + \frac{\beta_n}{\sqrt{2
\Omega^{out}_n}} e^{+i \Omega^{out}_n \eta}\,,
\end{equation}
with $|\beta_n|^2$ being the Bogolubov coefficient equal to the
particle creation number per mode $n$, and $\Omega^{out}_n \simeq
\sqrt{\epsilon_1} n$. Using the linear transformation
properties of hypergeometric functions,
we find that
\begin{eqnarray}
\left| \frac{\beta_n}{\alpha_n} \right| &=& e^{-2 \pi 
\tilde{b}} \left|\frac{ \cosh \pi(\tilde{d} + \tilde{b})}{ \cosh
\pi(\tilde{d} - \tilde{b})} \right| \label{betak}\,,
\end{eqnarray}
where $\tilde{b}$ and $\tilde{d}$ are given in terms of the Epstein
parameters~\cite{mar} $\epsilon_i$.

The family of of our dispersion models, for the frequency $n_{eff}$ ,
Fig. 1, attenuates to zero in the transplanckian regime
($\epsilon_2=0$), thereby
producing ultralow frequencies at very short  distances.  
Details about the exact solutions to the mode equations and the
spectrum computed through the method of Bogolubov coefficients can be
found in Ref.~\cite{mar}. The resulting
CMBR spectrum is shown to be (nearly) that of a {\em black body},
i.e., (nearly) scale invariant. Below we concentrate on the energy
contained in the ultralow frequencies range of modes, and show that it
is of the same order today as the matter energy density. 

\begin{figure}[t]
\epsfxsize=10cm
\epsfxsize=10cm
\hfil \epsfbox{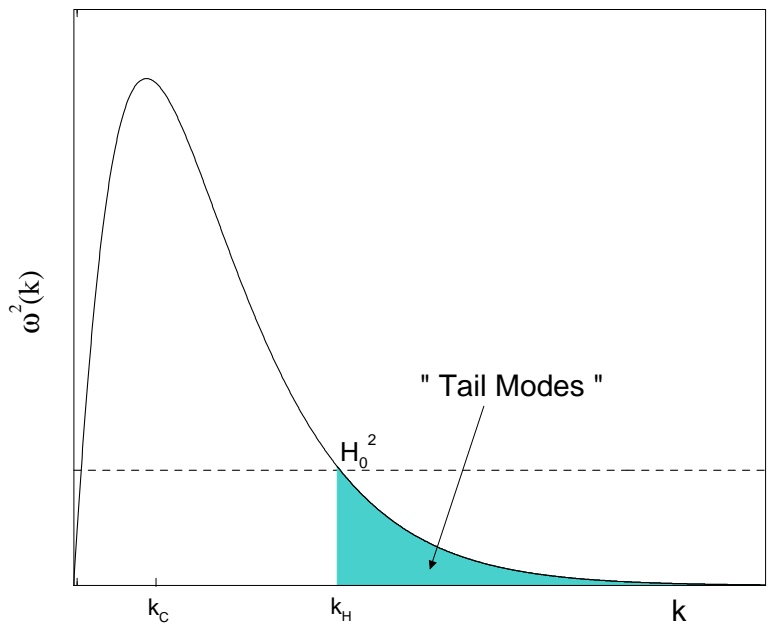} \hfil
\caption{{ Dispersion function and the range of modes in the tail, $k_H < k <
\infty$. $H_0$ is the present value of
the Hubble constant.} }
\label{fig2}
\end{figure}

For the special features of our choice of dispersion
relation Eq. (\ref{neff}), it is straightforward to calculate: i) which of the transplanckian modes have always been frozen to the present time and ii) the energy density contained in these modes. The modes at very high momenta but of ultra low frequencies 
$\omega(k)$ are frozen for as long as the Hubble expansion rate $H$ of
the Universe dominates over their frequency. We refer to that as the 
$tail$ of the dispersion graph. In Fig. (\ref{fig2}), for the
dispersed $\omega^2(k)$ vs. $k$, the tail corresponds to all the modes
beyond the point\footnote{It should be noticed that we have an
infinite ``tail'' of modes, with $k_H<k<\infty$.}  $k_H$, where $k_H$
is derived by the condition   
 $\omega^2(k_H)=H^2_0$, and $H_0$ is the Hubble rate
today. It then follows that they have not 
decayed and redshifted away but are still frozen $today$.  On the
other hand, since $H$
has been a decreasing function of time, many modes, even those in the
ultralow frequency range, have become dynamic and redshifted away one
by one, every time the expansion rate $H$ dropped below their
frequency. Clearly, these 
modes have long decayed into radiation\footnote{The reason why they 
behave as radiation can be traced back at their origin as metric
perturbations. Scalar perturbations produced during inflation do not
contribute to the total energy. Thus, these modes correspond to tensor
perturbations, i.e., gravitational waves of very short distance but
ultralow energy.}, and the tail modes are the only modes still frozen. They
contain vacuum energy of very short distance, hence of very low
energy. The last mode in the tail would decay when and if $H=0$. 
Their time-dependent behavior when they decay depends on the
evolution of $H$ and is complicated because they contribute to the
expansion rate for $H$. 

However we can calculate their contribution to the dark energy today,
when they are still frozen, thereby mimicking a cosmological
constant. The energy for the tail is given by:
\begin{equation}
\langle \rho_{tail} \rangle = \frac{1}{2 \pi^2 } \int^{\infty}_{k_H} k dk
\int \omega(k) d\omega~ |\beta_k|^2  \label{entail}\,,
\end{equation}
where the $|\beta_k|^2$ coefficient can be found in Ref.~\cite{mar}. 
We would also like to stress that due to
the dispersion class of functions chosen (defined in the whole range of
momenta from zero to infinity) the total
energy contribution of the modes produced is {\it finite}, and the
zero-point vacuum energy (the global cosmological cosntant) vanishes without the
need of applying any renormalization/subtraction scheme. In a sense, the
regularization-renormalization procedure is encoded in the class of 
dispersions we postulate. 

The numerical calculation of the tail energy produced the following
result: for random  different values of 
the free parameters, the dark energy of the tail is
$\rho_{tail}=10^{-122} f(\epsilon_1)$, times less than the total energy
{\em during inflation},
i.e. $\frac{\rho_{tail}}{\rho_{total}}=10^{-122} f(\epsilon_1)$ at 
Planck time. The prefactor $f(\epsilon_1)$, which depends weakly on the
parameter of the dispersion family $\epsilon_1$, is a small number
between 1 to 9, which clearly can contribute at the most by 1 order of
magnitude. 

This  $amazing$ result  can be understood qualitatively by noticing that the 
behavior of the frequency for the ``tail'' modes is nearly an
exponential decay (see Eq. (\ref{neff})), and as such dominates over the
other terms in the energy integrand of Eq. (\ref{entail}):
\begin{equation}
\omega^2(k > k_C) \approx exp(-k/k_C) \,,
\end{equation}  
Hence, due to the decaying exponential, the main contribution to the
energy integral in Eq.
(\ref{entail}) comes from the highest value of this exponentially
decaying frequency, which is the value of the integrand at the tail
starting point, $k_H \sim O(M_P)$, i.e., 
\begin{equation}
\langle \frac{\rho_{tail}}{\rho_{total}} \rangle \approx
\frac{k_H^2}{M_P^4} \omega^2(k_H) \approx \frac{H_0^2}{M_P^2}\approx
10^{-122} \,.  
\end{equation} 
Due to the physical requirement that the tail modes must have always been
frozen, the tail starting frequency $\omega(k_H)$ is then proportional to
the current value of Hubble rate $H_0$.  
We suspect that the above result of producing
such an extremely small number for the $tail$ energy {\em without any
fine-tuning } 
(and by using $M_{P}$ as the only fundamental scale of the theory), is
generic for any dispersion graph with a $tail$. 
First, all the modes with an
ultralow frequency $\omega<H_0$ will be frozen and thus produce dark
energy. Secondly, due to this kind of dispersion in the high momentum
regime, the phase space available for the ultra-low frequency modes with
$\omega(k>k_C)$ gets drastically reduced when compared to the phase
space factor of the degrees of freedom in the case of a non-dispersive transplanckian regime; 
controlling in this way these modes contribution to the energy density. 
Although the family of dispersion relations that feature a $tail$,
corresponding to vacuum modes of very short distances was motivated
by superstring duality~\cite{superstrings} it should be noted that the latter operates in the  target space while the former deals with a real spacetime. 

The tail equation of state, $w(t)= \langle p/\rho \rangle$, is an
$observable$ that will provide a test to the model~\cite{dark},
especially with the new data coming in the near future from the SNAP
\cite{snap} and SDSS~\cite{sdss} missions.  Although it
is straightforward  to calculate from the equations above, it is
technically difficult because of the strong coupling to the Friedmann
equations.   

The tail modes, that are frozen at present, provide a good
candidate for the dark energy as our calculations show. 
This idea is then a leap forward to this
longstanding and challenging problem of dark energy, for at least two 
reasons: first, inspired by superstring duality, it is very plausible
to speak of scenarios with ultralow frequencies and very high
momenta. Secondly, there was no fine-tuning involved or ad-hoc
parameters  in the
calculation for the dark energy, being a 122 orders of magnitude less
than the total energy during inflation.  


\end{document}